\documentclass[11pt]{article}

\usepackage[utf8]{inputenc}
\usepackage[T1]{fontenc}
\usepackage{amsmath, amssymb, amsthm}
\usepackage{mathtools}
\usepackage{bm}
\usepackage[margin=1in]{geometry}
\usepackage{graphicx}
\usepackage{booktabs}
\usepackage{array}
\usepackage{xcolor}
\usepackage{tcolorbox}
\usepackage{enumitem}
\usepackage{caption}
\usepackage{authblk}
\usepackage{hyperref}
\hypersetup{colorlinks=true, linkcolor=black, citecolor=black, urlcolor=blue}

\newtheorem{theorem}{Theorem}

\newtcolorbox{finding}[1][]{
  colback=orange!6, colframe=orange!55!black, boxrule=0.6pt,
  left=8pt, right=8pt, top=6pt, bottom=6pt, #1}

\newtcolorbox{authornote}[1][]{
  colback=blue!4, colframe=blue!40!black, boxrule=0.6pt,
  left=8pt, right=8pt, top=6pt, bottom=6pt, #1}

\newcommand{\R}{\mathbb{R}}
\newcommand{\Deff}{D_{\mathrm{eff}}}
\newcommand{\dmax}{d_{\max}}
\newcommand{\Ahat}{\hat{\bm{A}}}
\newcommand{\norm}[1]{\left\lVert #1 \right\rVert}
\newcommand{\Fnorm}[1]{\left\lVert #1 \right\rVert_F}

\title{\textbf{Stimulus-Evoked Network Dynamics in Human Cortical Organoids:\\
From a Graph-Computational Framework to Repeated-Stimulation Depression}}

\author[1]{Esmaeil S. Nadimi}
\author[1]{Vinay C. Gogineni}
\author[1]{Jan-Matthias Braun}
\author[2]{Martin Røssel Larsen}
\author[1]{Victoria Blanes-Vidal}
\author[2]{Helle Bogetofte Barnkob}
\affil[1]{[Applied AI and Data Science Unit, Maersk Mc-Kinney Moller Institute, Faculty of Engineering, University of Southern Denmark]}
\affil[2]{[Biomedical Mass Spectrometry and System Biology Unit, Department of Biochemistry and Molecular Biology, Faculty of Science, University of Southern Denmark]}
\date{}

\begin{document}
\maketitle

\begin{abstract}
\noindent
Human cortical organoids provide an experimentally accessible model of early neural
circuit formation, yet whether their activity reflects structured information processing
rather than spontaneous synchronization is unclear. We developed a graph-computational
framework to quantify stimulus-evoked propagation. This includes stimulus-conditioned functional
graphs, a graph-constrained dynamical (graph-neural-network) model used as a
system-identification tool, a biological message-passing principle bounding integration
depth by observable propagation depth ($\Deff \ge \dmax$), and a suite of graph-level
metrics. We carried this program out in full on longitudinal HD-MEA recordings from
three organoids. Once the true acquisition sampling rate and stimulus timing were
recovered, the evoked response proved to be a fast, near-synchronous network burst with
no measurable outward propagation (peak-latency vs.\ distance slope $\approx 0$). The
propagation/integration-depth metrics ($\Deff$, reachability index, $\dmax$) therefore do
not apply, and per-day connectivity graphs were not reliably estimable at the available
trial count, a negative result with methodological consequences for applying such
metrics to organoid data. Reframing around synchrony, response-population size and shared
variability revealed a control-validated phenomenon, i.e., repeated daily stimulation
progressively depressed and spatially contracted the evoked response. That repeated
stimulation reshapes organoid networks is established, but longitudinal designs in which
every preparation is stimulated cannot separate this from developmental maturation. We
break that confound with a developmentally-matched, stimulation-na\"ive control, where at day~7,
an organoid receiving its first-ever stimulation engaged ${\sim}93\%$ of the array,
whereas organoids with five prior sessions engaged ${\sim}10\%$. We further dissociate a
preserved first-trial response capacity from a degrading within-session endurance, and
identify independent within-session desynchronization.
\end{abstract}

\section{Introduction}

\subsection{Biological neural networks as perturbation-accessible computational systems}
The human cerebral cortex performs adaptive computation through distributed, recurrent
neural circuits operating under strict energetic and developmental constraints. Biological
neural networks self-organize through development, integrate information via local synaptic
interactions, and adapt through activity-dependent plasticity ~\cite{AlamElDin2025}. Human cortical organoids
provide an experimentally accessible model of developing neural tissue that recapitulates
key features of early corticogenesis, including excitatory--inhibitory diversity, radial
organization, and spontaneous network activity ~\cite{Trujillo2019,Sharf2022}. Whether organoids exhibit structured
distributed computation beyond spontaneous synchrony remains unclear ~\cite{Smirnova2023,Smirnova2024,Cai2023,BrownVarghese2026,Kagan2022,Robbins2026}. High-density
microelectrode arrays (HD-MEAs) enable simultaneous focal stimulation and large-scale
recording, providing a perturbation-based window onto how activity spreads through recurrent
circuitry following localized input.

\subsection{A graph-computational framework for evoked propagation}
\label{sec:framework}
We represent the recorded network as a graph $G=(V,E,\bm{A})$, where nodes $V$ are recorded
units (electrode-level multi-unit channels), edges $E$ are effective interactions, and
$\bm{A}$ is the weighted adjacency matrix \cite{BullmoreSporns2009}. Under stimulation, activity propagates along edges
through successive local integrations, so stimulus-evoked dynamics can be read as
graph-constrained signal transmission ~\cite{BullmoreSporns2009}. Graph topology then constrains computation along
principled axes, such as integration depth, reachability/controllability, encoding diversity,
dynamical dimensionality, and plasticity-induced reconfiguration ~\cite{Gilmer2017}. We formalize each of these
in Section~\ref{sec:metrics}.

\subsubsection{Biological message-passing theorem}
Inspired by Shannon's channel capacity, defined by the maximum rate of reliable transmission over a
channel and obtained by maximizing the mutual information $I(\cdot;\cdot)$ between input and
output ~\cite{Shannon1948}, we propose:

\begin{theorem}[Biological message passing]
\label{thm:bmp}
The effective integration depth required to predict stimulus-evoked responses in a biological
neural network is bounded below by the minimum number of graph hops separating the stimulation
site from the furthest reliably activated node. Formally, if stimulus information $S$ is
preserved at node $v$, i.e.\ the channel capacity satisfies
\begin{equation}
I(S; X_v) \;=\; \max_{p(s)} \; \big[\, H(X_v) - H(X_v \mid S)\,\big] \;>\; 0,
\label{eq:capacity}
\end{equation}
and $v$ lies $k = \mathrm{dist}_G(v_s, v)$ hops from the stimulation node $v_s$, then the
network supports at least $k$-hop information propagation, and consequently
\begin{equation}
\Deff \;\ge\; \dmax \;=\; \max_{v \in V_{\mathrm{active}}} \mathrm{dist}_G(v_s, v),
\label{eq:bmp}
\end{equation}
where $\Deff$ is the minimal message-passing depth reproducing the evoked dynamics
(Section~\ref{sec:gnn}) and $V_{\mathrm{active}}$ represents the set of reliably activated nodes.
\end{theorem}

\noindent Equation~\eqref{eq:bmp} is the testable core of the framework, as it predicts that the
model depth needed to reproduce responses grows with the spatial reach of evoked activity.
Section~\ref{sec:prop} reports the direct empirical test of its premise.

\subsection{Longitudinal stimulation and the direction of plasticity}
A complementary aim of this study was longitudinal, i.e., repeated daily stimulation could potentiate the
evoked response, depress it (habituation / synaptic depression), or track developmental maturation ~\cite{Osaki2024,Chow2025preprint,AlamElDin2025}. Distinguishing these requires both a longitudinal design and a control that
separates stimulation history from age. We therefore stimulated two organoids on days
$1,2,3,4,7$ and included a third recorded spontaneously at day~1 and stimulated for the first
time only at day~7, resulting in a single-stimulation, developmentally-matched control. This allowed us to study synchronous-burst regime
(participation-ratio dimensionality; the spirit of the reconfiguration index) and augment them
with measures appropriate to the observed response, i.e., magnitude, synchrony, responsive-population
size and spatial extent, and population coherence (Section~\ref{sec:readouts}).

\section{Methods}

\subsection{Preparations, stimulation protocol, and recordings}
Evoked activity was recorded from three cortical organoids (HD-MEA identifiers
$552, 613, 612$) on a 3Brain HD-MEA with $4096$ electrodes in a $64\times 64$ grid ~\cite{Berdondini2009,Amin2016,Schroter2025}; electrode
identity follows $\mathrm{id} = 64\,r + c$ for row $r$ and column $c$ (zero-indexed).
Stimulation was delivered through the array as a vertical line of electrodes (one column over a
contiguous row range). Each session comprised $N_{\mathrm{stim}}=10$ biphasic stimuli (nominal
$100~\mu$A, $400~\mu$s/phase) at a fixed inter-stimulus interval $\Delta = 30$~s. Stimulator and
recording shared the same hardware clock.

\begin{table}[h]
\centering
\caption{Stimulation geometry per organoid.}
\begin{tabular}{lcccc}
\toprule
Organoid & Stim column & Stim rows & \# electrodes & Electrode IDs \\
\midrule
552 & 29 & 10--38 & 29 & 669--2461 \\
613 & 13 & 17--42 & 26 & 1101--2701 \\
612 & 13 & 17--42 & 26 & 1101--2701 \\
\bottomrule
\end{tabular}
\end{table}

\noindent\textbf{Longitudinal design.} Organoids $552$ and $613$ were stimulated on days
$1,2,3,4,7$ (repeated). Organoid $612$ was recorded spontaneously at day~1 and stimulated only
at day~7 (single-stimulation control), sharing geometry with $613$.

\subsection{Sampling-rate correction and data-driven stimulus-time recovery}
\label{sec:timing}
Recordings were stored in BrainWave~6 \texttt{.bxr} format. The true acquisition rate
$f_s = 19{,}753.775$~Hz was read per file from the metadata field
\texttt{TimeConverter.FrameRate}; an incorrect default rate rescales spike times by a factor
$f_s/f_s^{\mathrm{default}}$ and progressively misaligns them from stimulus onsets (by up to
tens of seconds over a six-minute recording). Recordings were ${\sim}360$~s; two day-4
recordings ended earlier ($300$--$306$~s).

\noindent The files contained no explicit stimulus markers, and the nominal stimulation start varied
between sessions. We therefore recovered stimulus onsets from the data. Let $\rho(\tau)$ be the
population firing-rate trace (bin width $50$~ms). Candidate bursts are the supra-threshold
events of $\rho$, with a robust threshold
\begin{equation}
\theta \;=\; \mathrm{median}(\rho) \;+\; 5 \cdot 1.4826 \cdot \mathrm{MAD}(\rho).
\label{eq:thresh}
\end{equation}
where MAD stands for Median Absolute Deviation.
Given the known protocol (ten events spaced $\Delta$), we fit a periodic grid
$\{t_0 + (j-1)\Delta\}_{j=1}^{10}$ by selecting the start $t_0$ that maximizes the total matched strength (peak rate) over grid positions, assigning to each position the strongest
candidate within a tolerance window and rejecting weaker spontaneous bursts and edge artifacts:
\begin{equation}
t_0^\star \;=\; \arg\max_{t_0}\;\sum_{j=1}^{10}\;
\max_{\,b\,:\,|\tau_b - (t_0+(j-1)\Delta)|\le \delta}\; \rho(\tau_b),
\qquad \delta = 0.25\,\Delta .
\label{eq:grid}
\end{equation}
Each evoked onset $t_j$ is the local rise of its matched burst (walk-back from the rate peak to
the threshold crossing). This procedure was validated against trial-by-trial inspection.

\subsection{Stimulus-conditioned functional graph construction}
\label{sec:graphconstruct}
For each session we constructed a directed functional dynamic graph as a sequence of snapshots
\begin{equation}
G_t = (V_t, E_t, \bm{X}_t, \bm{A}_t),
\label{eq:dyngraph}
\end{equation}
with node-feature matrices $\bm{X}_t \in \R^{N\times F}$ (binned spike counts, $F=1$ in the
primary analysis; bin width $\mathrm{10}$~ms) and adjacency $\bm{A}_t \in \R^{N\times N}$. An edge
$j \to i$ was included when activity at $j$ reliably preceded activation at $i$ within the
post-stimulus window, exceeding a permutation threshold ($p<0.05$, False-Discovery-Rate (FDR)-corrected), with weight
proportional to the normalized evoked cross-correlation ~\cite{Poli2015,Sharf2022}:
\begin{equation}
A_{ij} \;\propto\; \max_{\Delta t > 0}\;
\frac{\big(x_j \star x_i\big)(\Delta t)}{\sqrt{(x_j\star x_j)(0)\,(x_i\star x_i)(0)}},
\label{eq:edge}
\end{equation}
where $\star$ denotes cross-correlation over the evoked window and $\Delta t>0$ enforces
$j$-before-$i$ precedence. The adjacency was symmetrically normalized for message passing ~\cite{KipfWelling2017},
\begin{equation}
\Ahat \;=\; \bm{D}^{-1/2}\,(\bm{A}+\bm{I})\,\bm{D}^{-1/2},
\qquad D_{ii} = \sum_j (\bm{A}+\bm{I})_{ij}.
\label{eq:norm}
\end{equation}
External stimulation is encoded as $\bm{S}_t \in \R^{N}$ with $S_{t,i}\neq 0$ only for
stimulation electrodes at stimulation bins. These graphs were constructed for every
organoid and day; where their cross-day analysis is reported in Section~\ref{sec:connresult}.

\subsection{Graph-constrained dynamical (GNN) model and effective depth}
\label{sec:gnn}
Stimulus-evoked dynamics were modeled as a discrete-time graph-structured system ~\cite{Gilmer2017},
\begin{equation}
\bm{X}_{t+1} \;=\; F_\theta(\bm{X}_t, \bm{S}_t; \Ahat)
\;=\; \sigma\!\Big( \Ahat\,\bm{X}_t\,\bm{W}_{\mathrm{neigh}}
\;+\; \bm{S}_t\,\bm{W}_{\mathrm{stim}} \Big),
\label{eq:gnn}
\end{equation}
with learnable $\bm{W}_{\mathrm{neigh}}\in\R^{F\times F}$, $\bm{W}_{\mathrm{stim}}\in\R^{1\times F}$,
and $\sigma=\mathrm{ReLU}$ (Xavier initialization; Adam, learning rate $\mathrm{1.0e-3}$, weight decay $\mathrm{1.0e-4}$, with cosine-annealing schedule). For a model
with $k$ stacked aggregation steps per time step, $F_\theta^{(k)}$ applies $k$ successive
neighborhood aggregations, and the trajectory error over $T$ post-stimulus bins is
\begin{equation}
\mathcal{L}_k \;=\; \frac{1}{T}\sum_{t=1}^{T}
\norm{\bm{X}_t^{\mathrm{pred}} - \bm{X}_t^{\mathrm{obs}}}_2^2 .
\label{eq:loss}
\end{equation}
The effective message-passing depth is
\begin{equation}
\Deff \;=\; \arg\min_{k}\; \mathcal{L}_k ,
\label{eq:deff}
\end{equation}
cross-validated across trials and stimulation sites. Data were split $70/15/15\%$
(train/validation/test); empirical, degree-preserving randomized, fully-connected, and
non-graph recurrent baselines were compared by paired permutation tests.

\subsection{Graph-computational metric suite (as evaluated)}
\label{sec:metrics}
We define the full metric suite, while Section~\ref{sec:metricfate} presents the empirical status of
each metric, given the findings.

\paragraph{Effective integration capacity (EIC).}
\begin{equation}
\mathrm{EIC} \;=\; \Deff .
\label{eq:eic}
\end{equation}
$\mathrm{EIC}=1$ indicates localized activation proximal to the stimulation site; while $\mathrm{EIC}>1$
indicates multi-hop recruitment.

\paragraph{Reachability index (RI).} With $R(v_s)=\{v\in V : \mathrm{dist}_G(v_s,v)<\infty\}$,
\begin{equation}
\mathrm{RI} \;=\; \frac{|R(v_s)|}{N} \;\in\; [0,1],
\label{eq:ri}
\end{equation}
the fraction of nodes reachable from the stimulation site.

\paragraph{Maximal propagation distance and BMP test.}
\begin{equation}
\dmax = \max_{v\in V_{\mathrm{active}}}\mathrm{dist}_G(v_s,v),
\qquad \text{test: } \Deff \ge \dmax \;\;\text{(Eq.~\eqref{eq:bmp})}.
\label{eq:dmaxtest}
\end{equation}

\paragraph{Stimulus separability index (SSI).} For stimulus $s$ with population response
vectors $\bm{x}_{s,k}\in\R^N$ over trials $k$, mean $\bar{\bm{x}}_s$,
\begin{equation}
d(s_1,s_2) = \norm{\bar{\bm{x}}_{s_1}-\bar{\bm{x}}_{s_2}}_2,
\qquad
\sigma_s^2 = \frac{1}{K_s}\sum_{k=1}^{K_s}\norm{\bm{x}_{s,k}-\bar{\bm{x}}_s}_2^2,
\qquad
\mathrm{SSI} = \frac{\mathbb{E}[\,d(s_1,s_2)\,]}{\mathbb{E}[\,\sigma_s\,]}.
\label{eq:ssi}
\end{equation}

\paragraph{Dynamical dimensionality (participation ratio, PR).} With response covariance
\begin{equation}
\bm{C} = \frac{1}{T}\sum_{t}(\bm{X}_t-\bar{\bm{X}})(\bm{X}_t-\bar{\bm{X}})^\top,
\qquad
\mathrm{PR} = \frac{\big(\sum_i \lambda_i\big)^2}{\sum_i \lambda_i^2},
\label{eq:pr}
\end{equation}
where $\{\lambda_i\}$ are the eigenvalues of $\bm{C}$. $\mathrm{PR}\to 1$ for a single global mode
(synchronized bursting), while larger PR indicates variance spread across independent modes ~\cite{LitwinKumar2017,Gao2017}.

\paragraph{Computational reconfiguration index (CRI).} For epochs $e_1,e_2$ with adjacency
$\bm{A}_{e}$, integration $\Deff^{(e)}$, and dimensionality $\mathrm{PR}^{(e)}$,
\begin{equation}
\Delta\bm{A} = \Fnorm{\bm{A}_{e_2}-\bm{A}_{e_1}},
\quad
\mathrm{CRI} = \alpha\,\Delta\bm{A} + \beta\,\big|\Deff^{(e_2)}-\Deff^{(e_1)}\big|
+ \gamma\,\big|\mathrm{PR}^{(e_2)}-\mathrm{PR}^{(e_1)}\big|,
\label{eq:cri}
\end{equation}
with nonnegative weights $\alpha,\beta,\gamma$.

\subsection{Reframed evoked-response read-outs}
\label{sec:readouts}
Let the response window be $W_r=[0,200]$~ms and baseline $W_b=[-250,-50]$~ms (equal widths,
$|W_r|=|W_b|$). For trial $k$ and electrode $i$, let $R_{ik}$ and $B_{ik}$ be the evoked and
baseline spike counts. We define three independent per-trial read-outs ~\cite{Mazzucato2016}.

\paragraph{Magnitude.}
\begin{equation}
M_k \;=\; \sum_{i=1}^{N} R_{ik} \;-\; \frac{|W_r|}{|W_b|}\sum_{i=1}^{N} B_{ik}.
\label{eq:mag}
\end{equation}

\paragraph{Synchrony.} Let $\{\tau_m\}$ be the evoked spike times (relative to onset) in $W_r$
on trial $k$. With temporal dispersion $\varsigma_k=\mathrm{SD}(\{\tau_m\})$,
\begin{equation}
\mathcal{S}_k \;=\; \frac{1}{\varsigma_k},
\label{eq:sync}
\end{equation}
defined only when the window contains at least $50$ spikes (else undefined). In this context, smaller dispersion meant 
 higher synchrony.

\paragraph{Dimensionality (per-trial participation ratio).} With response vector
$\bm{r}_k=(R_{1k},\dots,R_{Nk})$,
\begin{equation}
\mathcal{D}_k \;=\; \frac{\big(\sum_i R_{ik}\big)^2}{\sum_i R_{ik}^2},
\label{eq:dim}
\end{equation}
the effective number of contributing electrodes (a per-trial PR, Eq.~\eqref{eq:pr}).

\paragraph{Responsive electrodes (adaptive).} Electrode $i$ is responsive on a given day if its
evoked count reliably exceeds baseline across $K$ trials. With $d_{ik}=R_{ik}-B_{ik}$, mean
$\bar d_i$, SD $s_i$, the one-sided paired statistic is
\begin{equation}
t_i \;=\; \frac{\bar d_i}{s_i/\sqrt{K}},
\qquad
p_i = 1 - \mathcal{T}_{K-1}(t_i),
\label{eq:respt}
\end{equation}
with $\mathcal{T}_{K-1}$ the Student-$t$ CDF; $i$ is responsive if the Benjamini--Hochberg
FDR-adjusted $p_i < q$ ($q=0.05$) and $\bar d_i>0$ ~\cite{BenjaminiHochberg1995}. The responsive set is
$\mathcal{V}_{\mathrm{resp}}$, of size $n_{\mathrm{resp}}=|\mathcal{V}_{\mathrm{resp}}|$.

\paragraph{Within-session trend.} For a per-trial read-out $y_k$ ($k=1,\dots,K$) the
within-session trend is the Theil--Sen slope ~\cite{Sen1968}
\begin{equation}
\hat\beta = \operatorname{median}_{k<l}\frac{y_l - y_k}{\,l-k\,},
\label{eq:theilsen}
\end{equation}
with significance from Spearman's rank correlation between $y_k$ and $k$ (validly detected
trials only). Negative $\hat\beta$ for magnitude $M_k$ or synchrony $\mathcal{S}_k$ indicates within-session
depression / desynchronization.

\paragraph{Detrending and population coherence (PC1).} To remove the shared within-session
trend before assessing co-variation, each electrode's response series is linearly detrended, i.e., 
with design $\bm{u}=(1,2,\dots,K)^\top$ centered as $\tilde{\bm{u}}=\bm{u}-\bar u$, the residual
for electrode $i$ is
\begin{equation}
\tilde R_{ik} \;=\; R_{ik} - \Big(\bar R_i + \hat s_i\,\tilde u_k\Big),
\qquad
\hat s_i = \frac{\sum_k \tilde u_k\,(R_{ik}-\bar R_i)}{\sum_k \tilde u_k^2}.
\label{eq:detrend}
\end{equation}
Population coherence is the fraction of variance on the first principal component of the
detrended responsive-population matrix $\tilde{\bm{R}}\in\R^{K\times n_{\mathrm{resp}}}$:
\begin{equation}
\mathrm{PC1} \;=\; \frac{\lambda_1}{\sum_{m}\lambda_m},
\qquad \lambda_m = \mathrm{eig}_m\!\Big(\tfrac{1}{n_{\mathrm{resp}}}\,\tilde{\bm{R}}\,\tilde{\bm{R}}^\top\Big).
\label{eq:pc1}
\end{equation}
Computing PC1 on detrended data ensures it reflects genuine shared trial-to-trial variability
rather than the shared depression trend. PC1 is a stable, low-trial-count substitute for an
edge-level connectivity graph.

\paragraph{Spatial extent.} For responsive positions presented by rows and columns  $\{(r_i,c_i)\}_{i\in\mathcal{V}_{\mathrm{resp}}}$,
the spatial spread is
\begin{equation}
\Sigma \;=\; \sqrt{\,\mathrm{Var}(r_i) + \mathrm{Var}(c_i)\,}.
\label{eq:spread}
\end{equation}

\subsection{Statistical and interpretive notes}
In this study, we deployed two repeatedly-stimulated organoids and one single-stimulation control. Findings are reported as
reproducible, control-supported observations rather than population-level statistics. Across-day
comparisons report both the first-trial (``fresh'') response and the session mean. Per-day
pairwise connectivity (edge) graphs were constructed and evaluated, but are not a primary result
because at $K=10$ trials the pairwise correlation estimates are not reliably thresholdable
(Section~\ref{sec:connresult}). Supplementary preprocessing included band-pass $300$--$6000$~Hz;
stimulus-artifact blanking; threshold detection at $-5\times$ baseline noise SD and optional spike
sorting with curation.

\section{Results}

\subsection{The evoked response is a near-synchronous network burst, not a propagating wave}
\label{sec:prop}
Each stimulus reliably evoked a large population burst. Across all sessions, the ten stimuli
produced ten regularly-spaced bursts (inter-burst interval $30.1$~s, SD~$\approx 0$~s),
confirming reliable locking once timing was recovered (Eq.~\eqref{eq:grid}). At millisecond
resolution the burst peaked within tens of milliseconds of onset.

\begin{figure}[h]
\includegraphics[width=8cm, height=6cm]{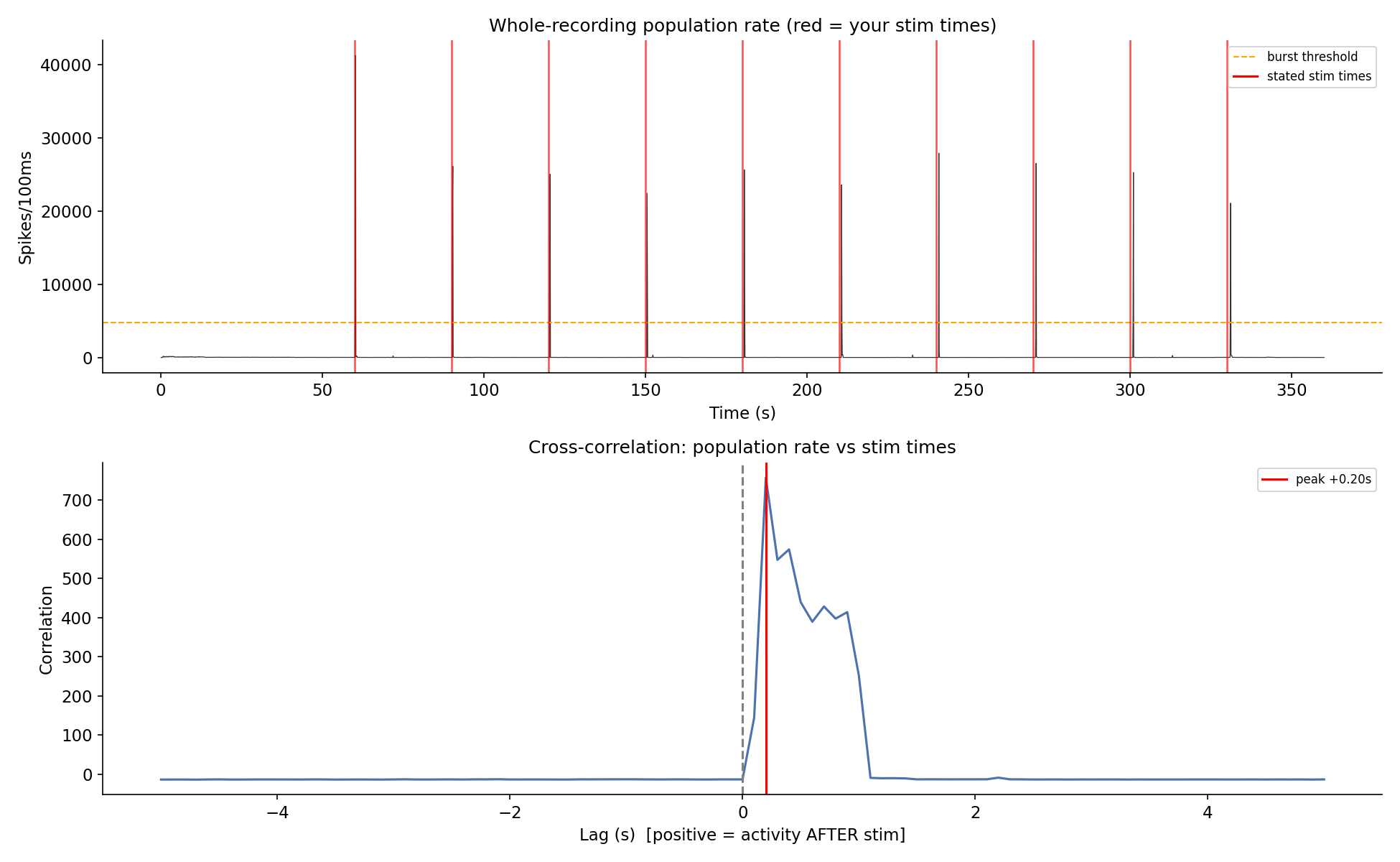}
\includegraphics[width=8cm, height=6cm]{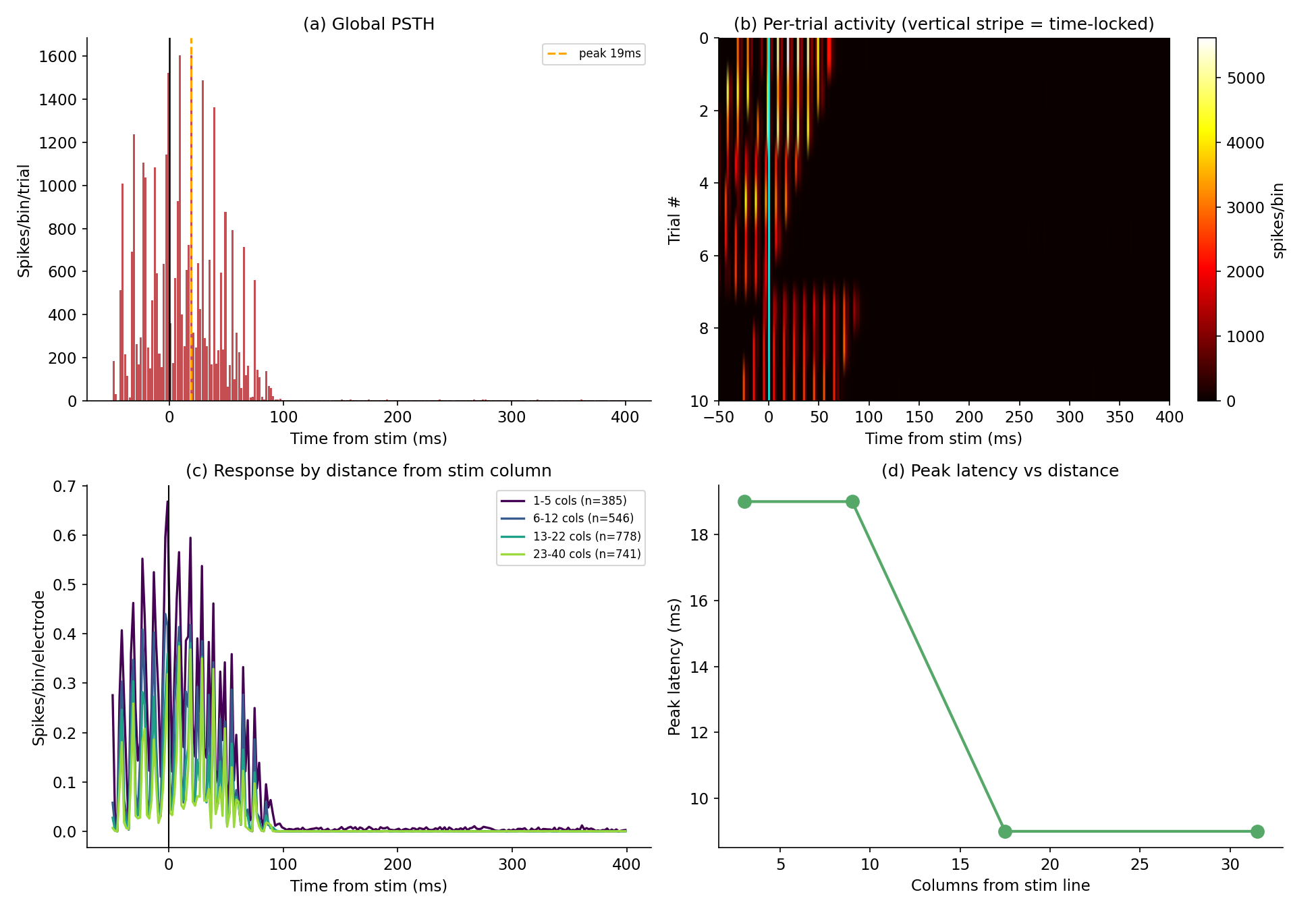}
\end{figure}

\noindent Direct test of Theorem~\ref{thm:bmp} indicated that grouping responses by distance from the
stimulation line, the peak-latency vs.\ distance slope was $\approx 0$~ms per electrode-column, meaning that the 
distal and proximal electrodes responded essentially simultaneously. The evoked response was a
near-synchronous, network-wide burst, not an outward-propagating wave. Hence $\dmax$
(Eq.~\eqref{eq:dmaxtest}) did not index a real spreading process, and $\mathrm{EIC}$
(Eq.~\eqref{eq:eic}), $\mathrm{RI}$ (Eq.~\eqref{eq:ri}), and the $\Deff\ge\dmax$ test were not
applicable to these recordings.

\noindent This is consistent with the dense recurrent connectivity expected in three-dimensional
cultures, which promotes rapid global synchronization. We proceed with the reframed read-outs
(Section~\ref{sec:readouts}).

\subsection{Repeated stimulation depresses the evoked response across days}
\label{sec:acrossday}
Across days, the overall evoked response of the repeatedly-stimulated organoids declined markedly
by day~7 (session-level response-to-baseline ratio):

\begin{table}[h]
\centering
\caption{Session-level evoked response strength (response/baseline ratio).}
\begin{tabular}{lccccc}
\toprule
Organoid & D1 & D2 & D3 & D4 & D7 \\
\midrule
552 (repeated) & 1123 & 1411 & 1802 & 1450 & \textbf{206} \\
613 (repeated) & 487 & 897 & 230 & 1128 & \textbf{268} \\
612 (single-stim control) & --- & --- & --- & --- & \textbf{1914} \\
\bottomrule
\end{tabular}
\end{table}

\noindent\textbf{The single-stimulation control isolates stimulation history from age.} At day~7
the repeatedly-stimulated organoids were strongly depressed (ratios $206$, $268$), whereas control
$612$, the same age but stimulation-na\"ive until day~7 produced the strongest response in the
dataset ($1914$), ${\sim}7$--$9\times$ larger. Because $612$ was developmentally matched, the day-7
depression in $552$ and $613$ is driven by repeated-stimulation history rather than maturation.

\begin{figure}[h]
\centering
\includegraphics[width=15cm, height=6cm]{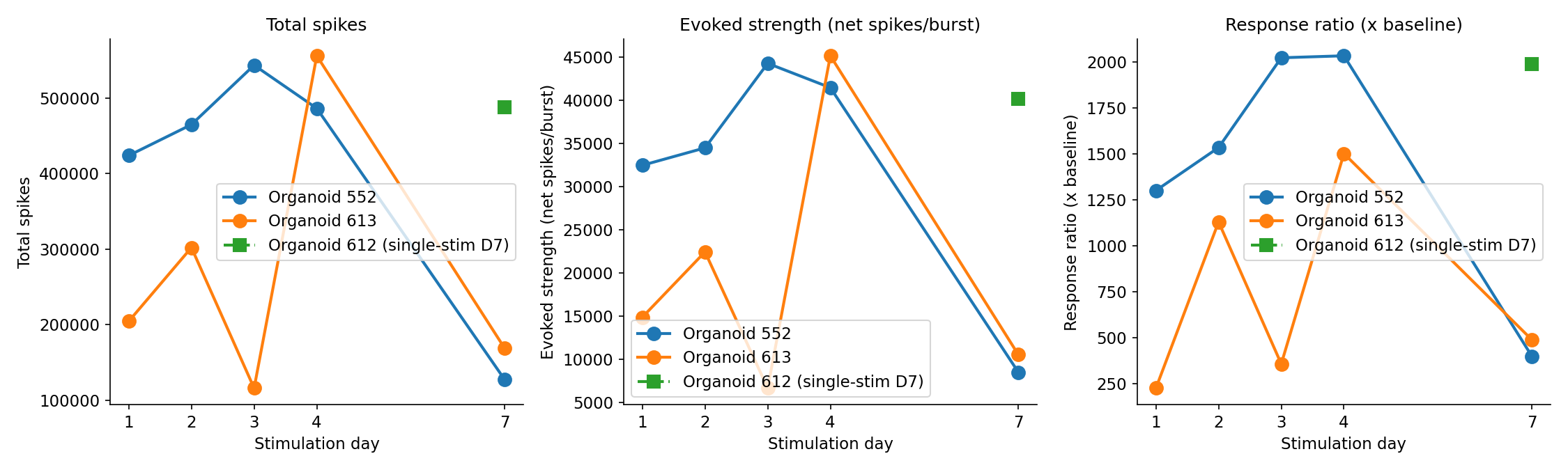}
\end{figure}

\noindent It is worth mentioning that Organoid $613$ was non-monotonic across days (low D3, high D4), unlike the
cleaner rise-then-fall of $552$. The depressed D7 endpoint is consistent across both, but the
intermediate trajectory is variable; and D3/D4 recordings warrant the scrutiny in
Section~\ref{sec:limits}.

\subsection{Across-day depression reflects loss of within-session endurance, not capacity}
\label{sec:capacity}
Separating the first-trial ($M_1$) from the session-mean response showed these behave very
differently. First-trial magnitude was relatively preserved $D1\to D7$ (552: ${\sim}50{,}000\to{\sim}37{,}000$
net spikes; 613: ${\sim}46{,}000\to{\sim}42{,}000$), whereas the session mean
$\bar M = \tfrac{1}{K}\sum_k M_k$ collapsed (552: ${\sim}32{,}000\to{\sim}8{,}500$;
613: ${\sim}15{,}000\to{\sim}10{,}500$). The ratio $M_1/\bar M$ grew from ${\sim}1.5$--$3\times$
early to ${\sim}6\times$ by day~7. The network's capacity to mount a strong initial daily
response was largely preserved; however, the ability to sustain responses across the
session was degraded. The control was consistent, as $612$ at day~7 had both the highest first-trial response
(${\sim}73{,}500$) and a high session mean.

\begin{figure}[h]
\centering
\includegraphics[width=15cm, height=6cm]{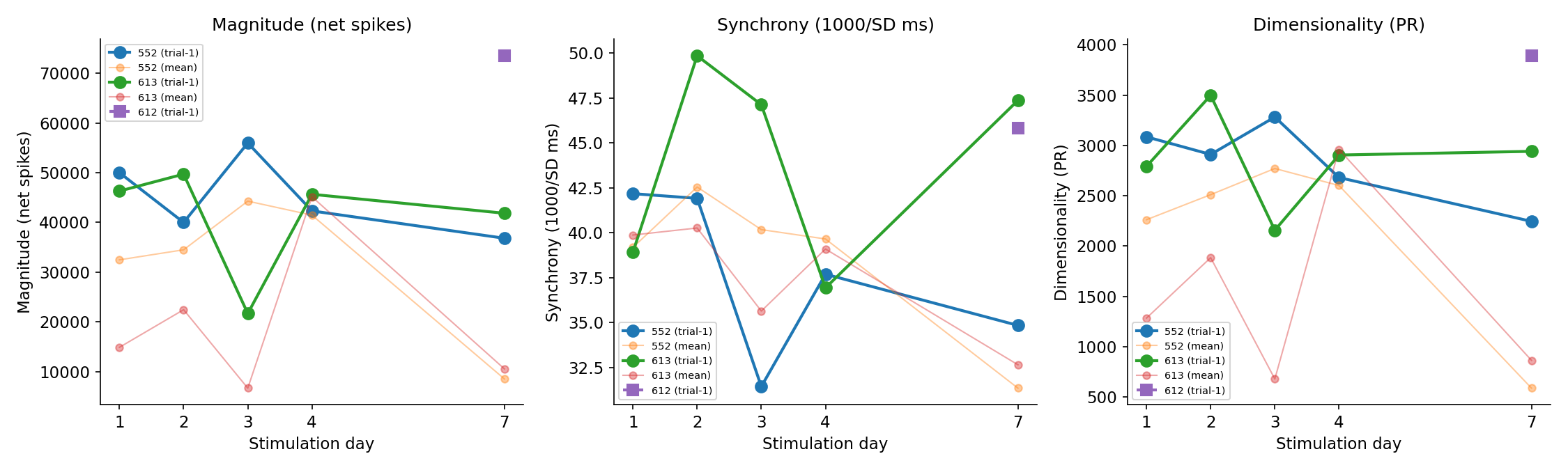}
\end{figure}

\subsection{Within-session dynamics: magnitude depression and independent desynchronization}
\label{sec:within}
Within sessions, the response typically depressed across the ten trials, most often a rapid drop
after the first stimulus, and then a lower plateau (synaptic-depression-like). Significant negative
within-session magnitude slopes (Eq.~\eqref{eq:theilsen}) occurred in multiple recordings, including
organoid $613$ on essentially every day and the control $612$ (steepest decline, despite no prior
stimulation). Organoid $552$ showed clear within-session depression on day~1 but a noisier pattern
otherwise.

\begin{figure}[h]
\centering
\includegraphics[width=15cm, height=6cm]{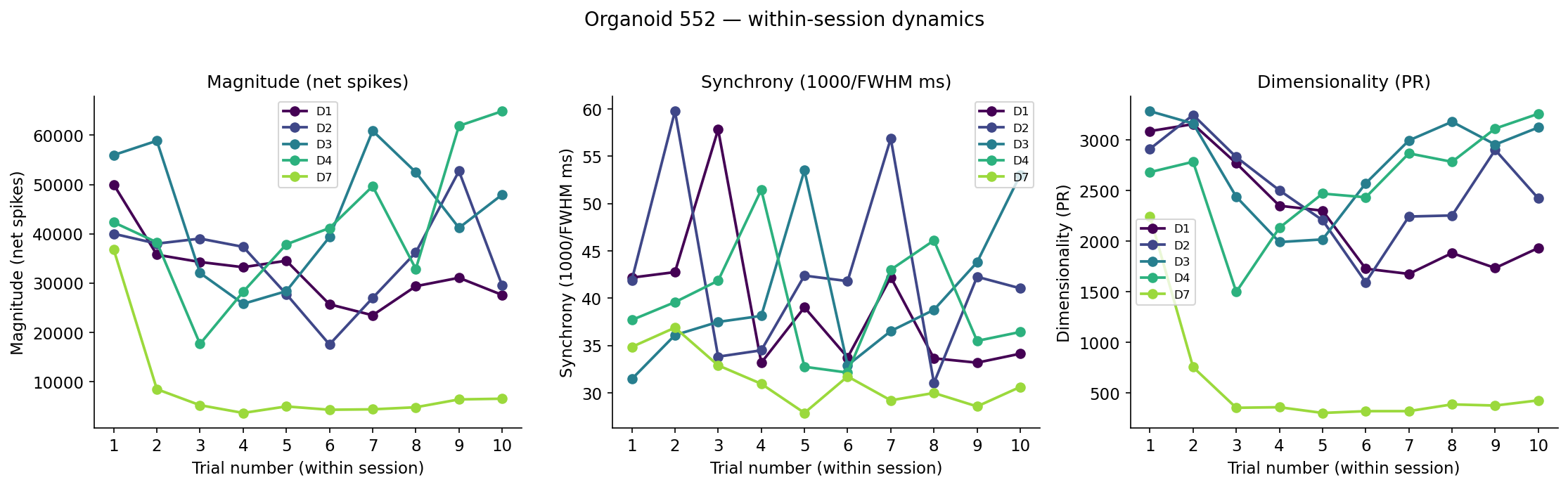}
\end{figure}

\noindent We further observed that synchrony depresses independently of magnitude. Treating $\mathcal{S}_k$
(Eq.~\eqref{eq:sync}) separately revealed a dissociation, i.e.,  in several recordings, the burst became
progressively less synchronous across trials, not always tracking magnitude (e.g.\ $552$
day~7 desynchronized significantly with no magnitude trend). The stimulation-na\"ive control $612$
showed strong magnitude depression but no significant desynchronization, whereas the
repeatedly-stimulated $613$ desynchronized, suggesting (hypothesis, given small $n$) that
desynchronization is more associated with prior stimulation while magnitude depression is intrinsic.

\noindent As the control shows clear within-session depression on its first-ever session, fast
within-session depression does not require prior stimulation, and as a result, the slow across-day component is what
accumulates with repeated daily stimulation, suggesting partly distinct mechanisms.

\subsection{The responding network spatially contracts with repeated stimulation}
\label{sec:contract}
The size of the responsive population $n_{\mathrm{resp}}$ (Eq.~\eqref{eq:respt}) showed the clearest,
most robust effect.

\begin{table}[h]
\centering
\caption{Responsive-electrode count and array fraction across days.}
\begin{tabular}{lccccc}
\toprule
Organoid & D1 & D2 & D3 & D4 & D7 \\
\midrule
552 responsive       & 3852 & 4020 & 4037 & 4005 & \textbf{384} \\
\quad(\% array)      & 94\% & 99\% & 99\% & 99\% & \textbf{9.5\%} \\
613 responsive       & 1711 & 2932 & 574  & 4048 & \textbf{486} \\
\quad(\% array)      & 42\% & 72\% & 14\% & 100\% & \textbf{12\%} \\
612 control responsive & --- & --- & --- & --- & \textbf{3788} \\
\quad(\% array)      & --- & --- & --- & --- & \textbf{93\%} \\
\bottomrule
\end{tabular}
\end{table}

\noindent In both repeatedly-stimulated organoids the responding population collapsed from
essentially the whole array to ${\sim}10\%$ by day~7 ($552$: $384$; $613$: $486$), whereas the
single-stimulation control responded on $93\%$ ($3788$ electrodes). At matched age, a first
stimulation engaged ${\sim}3800$ electrodes versus ${\sim}400$ after five prior sessions, a
${\sim}10$-fold difference. The surviving day-7 responders in $552$ were also spatially contracted
($\Sigma$ from ${\sim}26$ to ${\sim}10$, Eq.~\eqref{eq:spread}), indicating contraction toward a
smaller region rather than uniform thinning.

\begin{figure}[h]
\centering
\includegraphics[width=8cm, height=5cm]{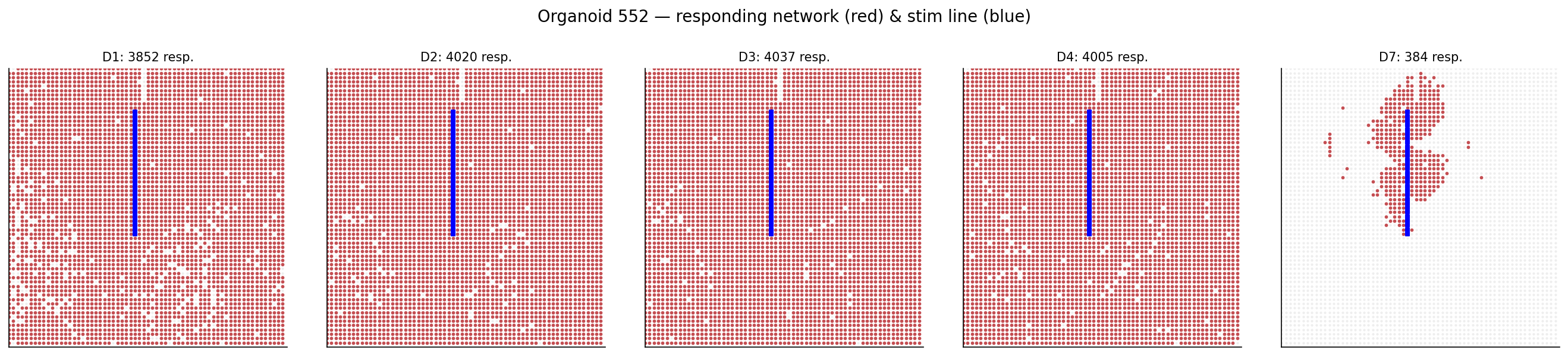}
\includegraphics[width=8cm, height=5cm]{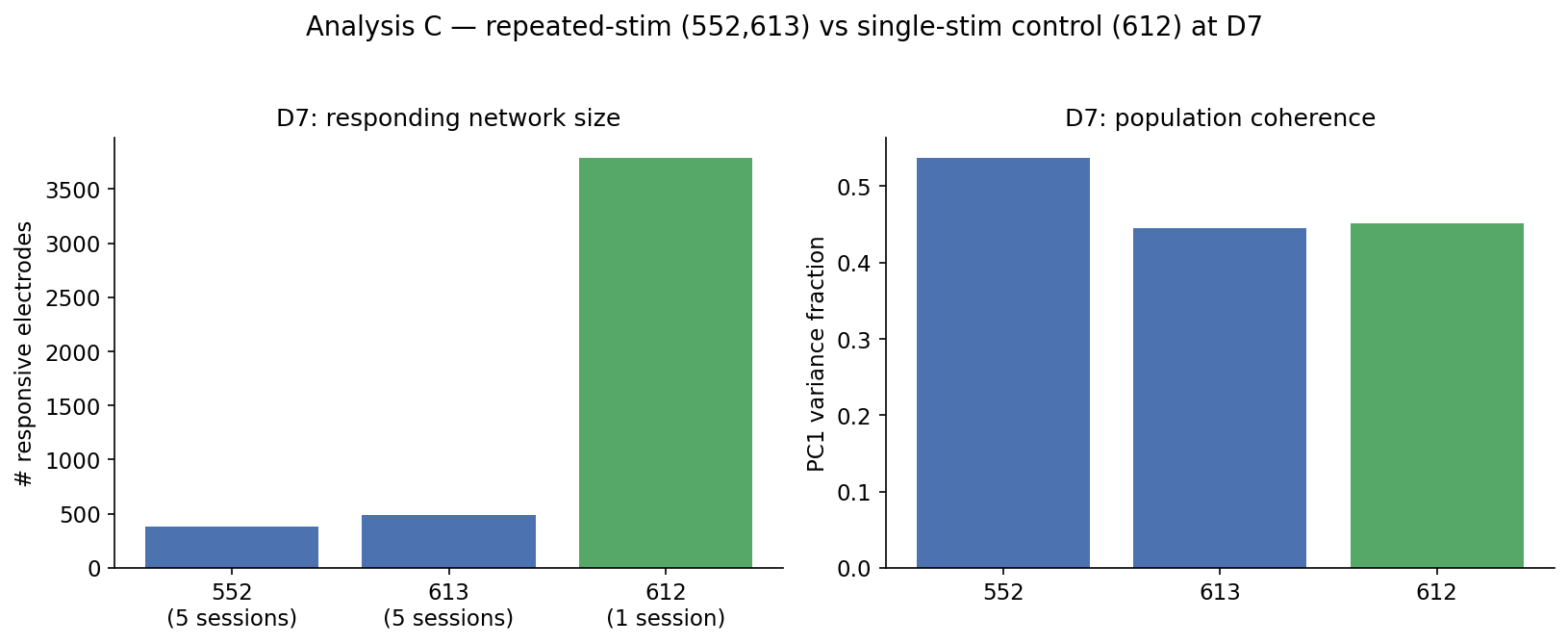}
\includegraphics[width=8cm, height=5cm]{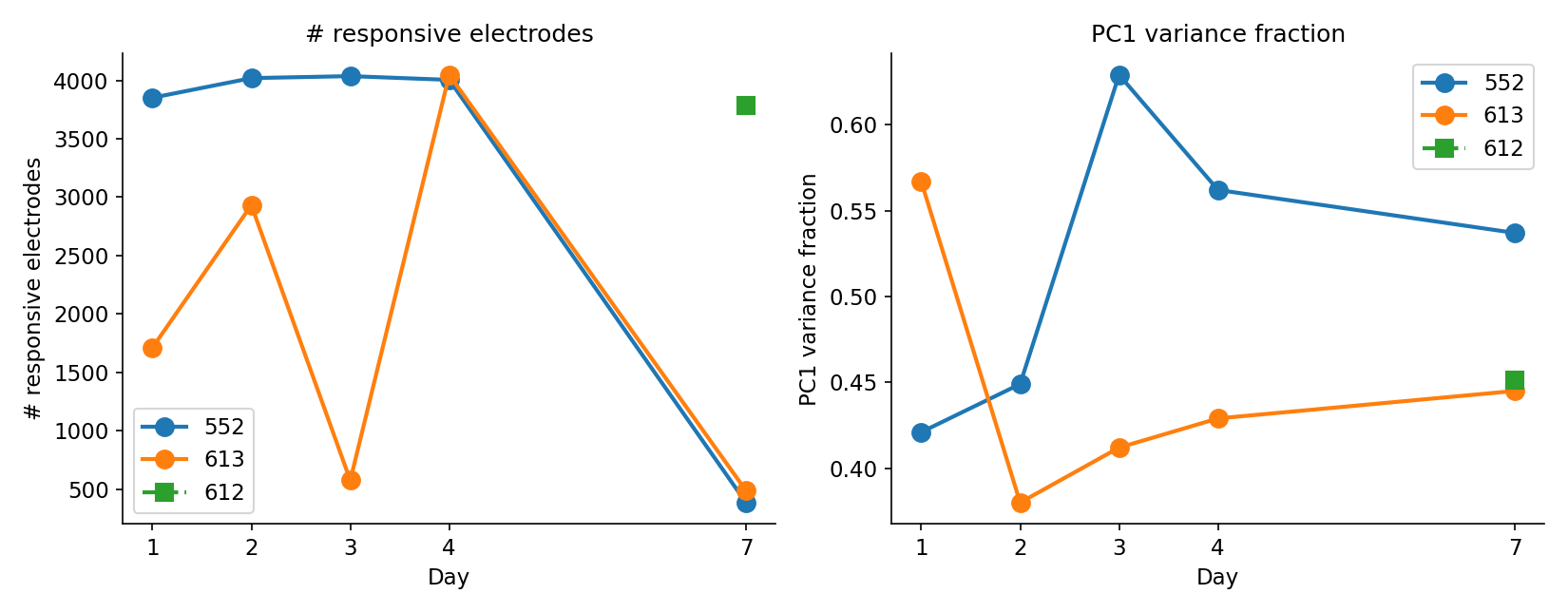}
\end{figure}

\noindent It is worth highlighting that the coherence was preserved (retained PR/PC1 metric). The detrended population coherence
$\mathrm{PC1}$ (Eq.~\eqref{eq:pc1}) stayed ${\sim}0.4$--$0.6$ across all conditions, including the
small day-7 networks. Repeated stimulation predominantly removed electrodes from the
responding pool rather than desynchronizing those that remain, such that the effect was on population
size more than coherence.

\subsection{Per-day connectivity graphs were evaluated but are not reliable at $K=10$}
\label{sec:connresult}
We constructed the per-day functional graphs of Section~\ref{sec:graphconstruct} and computed their
co-variation structure on the detrended responsive population, with edge significance from a
trial-shuffle permutation null. At $K=10$ trials (effectively $K-1$ after detrending) the null
distribution of $|r|$ is broad, i.e., permutation thresholds approached $|r|\approx 0.98$, yielding
near-empty thresholded graphs, and two detrending variants (linear detrend vs.\ drop-trial-1)
disagreed substantially on derived measures (density, modularity, largest component). We therefore
do not report edge-level graph statistics as primary results; and the robust network-level conclusions
rest on $n_{\mathrm{resp}}$ (Eq.~\eqref{eq:respt}) and $\mathrm{PC1}$ (Eq.~\eqref{eq:pc1}). This is a
limitation of trial count, not of the construction, as the graphs were built and evaluated for every
organoid and day.

\subsection{Data-quality considerations and limitations}
\label{sec:limits}
\begin{itemize}[leftmargin=1.4em]
\item \textbf{Day-3/day-4 anomalies.} Organoid $613$ showed an implausibly non-monotonic responsive
trajectory ($42\to72\to14\to100\to12\%$). Day~3 had the lowest total spike count; and both day-4
recordings were truncated ($8$--$9$ usable trials). These are under separate diagnostic examination
(activity level vs.\ threshold fragility vs.\ baseline elevation vs.\ truncation).
\item \textbf{Mis-aligned vs.\ failed trials.} Some apparent single-trial failures were detection
mis-alignments (a small spontaneous burst selected instead of the true evoked burst), but selecting the
dominant burst (Eq.~\eqref{eq:grid}) recovered the true responses.
\item \textbf{One silent stimulation electrode} in $613$ (ID $1421$) recorded no spikes.
\item \textbf{Low-trial limits.} Edge-level connectivity is not reliably estimable at ten trials
(Section~\ref{sec:connresult}).
\item \textbf{Control replication.} The single-stimulation control is one organoid, and additional
controls would strengthen the dissociation of stimulation history from development.
\end{itemize}

\subsection{Status of the original metric suite}
\label{sec:metricfate}
\begin{table}[h]
\centering
\caption{Status of each originally-proposed metric under the empirical findings}
\renewcommand{\arraystretch}{1.2}
\begin{tabular}{p{3.2cm} p{5.0cm} p{6.2cm}}
\toprule
Metric & Original role & Status in this study \\
\midrule
BMP test ($\Deff\!\ge\!\dmax$), Eq.~\eqref{eq:bmp} & Link propagation depth to integration & Set aside --- no propagation (slope $\approx 0$) \\
EIC $=\Deff$, Eq.~\eqref{eq:eic} & Multi-hop integration depth & Not applicable --- synchronous burst \\
RI, Eq.~\eqref{eq:ri} & Fraction of network reached & Not informative --- near-global, no structure \\
SSI, Eq.~\eqref{eq:ssi} & Stimulus-specific encoding & Not estimated --- single stimulation condition per organoid \\
PR, Eq.~\eqref{eq:pr} & Effective \# activity modes & \textbf{Retained} --- per-trial read-out \& PC1 coherence \\
CRI, Eq.~\eqref{eq:cri} & Plasticity across epochs & \textbf{Retained} --- across-day change in valid read-outs \\
\bottomrule
\end{tabular}
\end{table}

\section{Discussion}

We set out to test whether stimulus-evoked activity in human cortical organoids reflects
structured, multi-hop signal propagation that could be quantified with a graph-computational
framework, and to characterize how that activity is reshaped by repeated stimulation. Two results
frame everything that follows. First, once acquisition timing was correctly recovered, the evoked
response proved to be a fast, near-synchronous, network-wide burst rather than a spatially
propagating wave, so the propagation/integration-depth machinery we had developed
($\Deff \ge \dmax$, EIC, RI) does not apply to these preparations. Second, and independently,
repeated daily stimulation produced a reproducible, control-validated depression and spatial
contraction of the evoked response. We discuss each in turn, then their mechanistic
interpretation and limitations.

\subsection{Relation to prior work and the contribution of a stimulation-naïve control}

Repeated stimulation alters organoid network activity, as optogenetic entrainment of connected cerebral organoids induces sustained changes in later responses~\cite{Osaki2024}, and two weeks of daily
electrical stimulation on high-density arrays reshapes response patterns,
spontaneous activity and functional connectivity~\cite{Chow2025preprint}.
What such longitudinal designs cannot settle is "why". Because every
preparation is stimulated, the observed change is confounded with ordinary
developmental maturation over the same interval, and maturation is not a minor
alternative in organoid preparations, where network activity changes
substantially across days in culture~\cite{Trujillo2019}. Our design breaks
this confound directly. Organoid 612 was cultured alongside the others,
recorded spontaneously at day~1, and stimulated for the first time only at
day~7. At matched age it engaged 3788 electrodes ($93\%$ of the array), while
the two organoids with five prior sessions engaged 384 and 486 ($\approx\!10\%$), an order-of-magnitude difference attributable to stimulation history rather
than age. This control, rather than the depression itself, is the load-bearing
element of our claim.

\noindent Three further results follow from the reframed read-outs. First, the depression
is not a loss of capacity but of endurance. The first evoked response of
each day was largely preserved from day~1 to day~7 (552: ${\sim}50{,}000 \to
{\sim}37{,}000$ net spikes; 613: ${\sim}46{,}000 \to {\sim}42{,}000$), whereas
the session mean collapsed and the ratio $M_1/\bar{M}$ grew from
${\sim}1.5$--$3\times$ to ${\sim}6\times$. What repeated stimulation degrades is
the network's ability to sustain responses across a session, not to mount one.
Second, the effect is on population size rather than coherence. The
responding population contracted spatially ($\Sigma$ from ${\sim}26$ to
${\sim}10$) while its shared-variance structure was preserved
($\mathrm{PC1} \approx 0.4$--$0.6$ throughout), indicating that electrodes drop
out of the responsive pool rather than desynchronising within it. Third, we
report a negative result of methodological consequence. With stimulus timing
correctly recovered, the evoked response showed no latency--distance gradient,
so propagation-based graph metrics, such as reachability, integration depth, and
bounds of the $\Deff \ge \dmax$ form quantify a process these preparations do
not exhibit, and should be justified empirically before being applied.

\subsection{Why the evoked response is near-synchronous rather than propagating}
The absence of a latency--distance gradient (peak-latency slope $\approx 0$) indicates that focal
stimulation recruits the network essentially simultaneously rather than through sequential,
distance-dependent recruitment. This is consistent with what is known about maturing
three-dimensional neural cultures, in which dense recurrent excitatory connectivity and
developing inhibition give rise to network-wide synchronized bursting
\cite{Trujillo2019}.
In such a regime, once activity crosses a threshold anywhere in the recurrent network, it engages
the whole connected population within a few tens of milliseconds, so the observable is the
presence, size, and timing precision of a global burst rather than a traveling wavefront.
This has a direct methodological implication that generalizes beyond our preparations, i.e., graph
metrics premised on measurable multi-hop propagation, such as reachability depth, integration depth,
and any lower bound of the $\Deff \ge \dmax$ form are only meaningful when the evoked response
actually exhibits distance-dependent latency structure. We recommend that such structure be
verified empirically (as in our Section~3.1 propagation check) before propagation-based
graph metrics are applied to organoid or culture data; otherwise these metrics quantify a process
the tissue does not exhibit. We regard the transparent reporting of this negative result, alongside
the framework that motivated it, as a contribution in its own right

\subsection{Repeated stimulation depresses and spatially contracts the evoked response}
The central positive finding is that repeated daily stimulation progressively depressed the evoked
response, culminating in a striking collapse of the responding electrode population, from
essentially the whole array to ${\sim}10\%$ by day~7 in both repeatedly-stimulated organoids, while
a developmentally-matched organoid receiving its first-ever stimulation at day~7 retained
${\sim}93\%$. Because the control was the same age but stimulation-na\"ive, the collapse cannot be
attributed to maturation, viability decline, or array degradation over the recording period; it is
specifically associated with the history of repeated stimulation. Repeated stimulation has been shown to induce sustained changes in organoid network responses and functional connectivity~\cite{Osaki2024,Chow2025preprint}. Our results extend this in two ways, firstly by including a developmentally-matched, stimulation-naïve control that separates stimulation history from maturation, and secondly by dissociating a preserved first-trial response capacity from a progressively degrading within-session endurance. We showed that by including a stimulation-naïve, developmentally-matched control, the change is a progressive depression and spatial contraction of the responding population.
\cite{Cai2023,Kagan2022,Osaki2024}.

\subsection{Dissociating capacity from endurance: candidate mechanisms}
A key refinement is that the across-day depression did not reflect loss of the network's maximal
capacity. The first (``fresh'') evoked response of each day was relatively preserved from day~1 to
day~7, whereas the session-averaged response collapsed, because within-session depression became
progressively more severe on later days. In other words, repeated stimulation degrades the
network's ability to sustain responses across a session more than its ability to mount an initial one. Several non-exclusive mechanisms are consistent with this dissociation:

\begin{itemize}[leftmargin=1.4em]
\item \textbf{Short-term synaptic depression / vesicle-pool depletion.} The rapid within-session
drop after the first stimulus, followed by a lower plateau, is the classic signature of short-term
synaptic depression through depletion of the readily-releasable vesicle pool and its incomplete
recovery at the 30~s inter-stimulus interval
\cite{ZuckerRegehr2002,AlamElDin2025}.
That the stimulation-na\"ive control showed the steepest within-session depression on its
first-ever session indicates this fast component is intrinsic and history-independent.
\item \textbf{Slow, accumulating homeostatic or metabolic change.} The across-day component, the
worsening of within-session sustainability and the contraction of the responding pool accumulate
only with repeated daily stimulation. Candidates include homeostatic synaptic downscaling in
response to repeated strong drive
\cite{Turrigiano1998,TurrigianoNelson2004},
activity-dependent shifts in excitation--inhibition balance, and slower metabolic or
excitotoxic costs of repeated large-scale synchronous bursting.
\item \textbf{Spatial contraction without loss of coherence.} That the responding population shrinks
in size while its shared-variance coherence (PC1) is preserved suggests that repeated
stimulation progressively removes electrodes/units from the responsive pool rather than
degrading the synchrony of those that remain. This is more consistent with a loss of recruitable
periphery, i.e., units near threshold dropping out as excitability or synaptic efficacy declines than
with a global desynchronization of the core.
\end{itemize}

\subsection{Independent desynchronization and a possible signature of stimulation history}
Treating synchrony as a read-out separate from magnitude revealed a dissociation. In several
recordings, the evoked burst became progressively less temporally precise across trials,
independently of whether its magnitude declined. Notably, the stimulation-na\"ive control showed
strong magnitude depression but no significant desynchronization, whereas a repeatedly-stimulated
organoid desynchronized. We advance, as a hypothesis to be tested with larger samples, that
within-session magnitude depression is an intrinsic, history-independent property of these
networks, whereas progressive desynchronization is more associated with a prior history of
repeated stimulation, potentially reflecting an accumulating disruption of the inhibitory or
gap-junctional mechanisms that sharpen population timing
\cite{Trujillo2019}.
Given $n=1$ control, this remains a hypothesis rather than a conclusion.

\subsection{Methodological lessons}
Two methodological points have broad relevance for organoid HD-MEA studies. First, correct
recovery of the acquisition sampling rate and of stimulus timing was decisive. An incorrect default
rate rescaled spike times and misaligned them from stimulus onsets by up to tens of seconds,
which alone can manufacture or destroy apparent latency structure. Data-driven recovery of
stimulus onsets directly from the population-rate trace rather than reliance on nominal protocol
times proved essential given session-to-session variation in stimulation start. Second, the number
of stimulus repetitions imposes a hard statistical ceiling on what can be estimated. At ten trials,
pairwise functional-connectivity graphs over thousands of electrodes were not reliably thresholdable
(permutation thresholds approaching $|r|\approx0.98$), whereas aggregate measures, responsive-population
size and first-principal-component coherence were stable and interpretable. Studies intending to
estimate edge-level connectivity from evoked responses should plan trial counts accordingly.

\subsection{Limitations}
Our conclusions are tempered by several limitations, stated forthrightly. The design comprises two
repeatedly-stimulated organoids and a single single-stimulation control; findings are therefore
reproducible, control-supported observations rather than population-level statistical claims, and the
control-based dissociation of stimulation history from maturation rests on one control organoid.
The day-3 and day-4 recordings of one organoid seemed anomalous, as day-3 session had the lowest
total activity of the dataset and both day-4 recordings were truncated to eight or nine usable
trials, so the intermediate day-to-day trajectory of that organoid should be interpreted with
caution even though the day-7 endpoint was consistent across both repeatedly-stimulated organoids.
One stimulation electrode in one organoid recorded no spikes. Finally, because the evoked response
is a near-synchronous burst, our study does not, and cannot, adjudicate whether organoids support
structured multi-hop computation under other stimulation regimes. It establishes only that they do
not exhibit measurable propagation under the focal, single-site protocol used here.

\subsection{Future work}
Several directions would directly extend and strengthen these findings.

\begin{itemize}[leftmargin=1.4em]
\item \textbf{Increase sample size and add controls.} The most important next step is to replicate the
across-day depression and the single-stimulation control in a larger cohort of organoids, enabling
population-level statistics and a properly powered test of the stimulation-history-versus-maturation
dissociation, including multiple single-stimulation controls at several ages.

\item \textbf{Resolve recovery dynamics and reversibility.} Whether the day-7 depression recovers
after a stimulation-free interval, and on what timescale, would distinguish a transient homeostatic
adaptation from a durable reconfiguration ~\cite{Robbins2026}. A design interleaving stimulation and rest days, with
recordings during recovery, would address this.

\item \textbf{Vary the inter-stimulus interval and protocol.} Because within-session depression likely
reflects incomplete recovery between stimuli, systematically varying the inter-stimulus interval
(and total stimulus number) would map the recovery time-constant of the fast depression and test
whether longer intervals spare within-session endurance.

\item \textbf{Multi-site stimulation to test encoding and separability.} The stimulus separability index (SSI) requires at least two distinguishable stimulus conditions and could not be estimated here because each organoid received a single, fixed stimulation pattern (one electrode column, applied identically on every trial). Delivering two or more spatially distinct stimulation patterns, for example, columns at different array positions  interleaved across trials would allow a direct test of whether organoids produce separable population responses to different inputs, and whether such separability is itself degraded by repeated stimulation.

\item \textbf{Denser trial sampling for connectivity.} Increasing the number of stimulus repetitions
per session, at the cost of longer recordings or shorter intervals would raise the statistical
ceiling enough to make edge-level functional-connectivity graphs reliably estimable, allowing the
originally-intended graph-structural analyses (density, modularity, community structure) to be
revisited on firmer ground ~\cite{DuenkiIkeuchi2026}.

\item \textbf{Mechanistic and pharmacological dissection.} Targeted manipulations, blocking specific
receptors or transporters, or perturbing inhibition during the repeated-stimulation paradigm would
help adjudicate among the candidate mechanisms (vesicle-pool depletion, homeostatic downscaling,
excitation--inhibition shifts, metabolic cost) proposed above, and connect the network-level
phenomenology to cellular substrates.

\item \textbf{Cross-scale validation.} Complementary read-outs, such as calcium imaging for single-cell
spatial resolution, or molecular/transcriptomic profiling before and after repeated
stimulation would test whether the electrophysiological contraction of the responding population
corresponds to identifiable cellular or molecular changes.
\end{itemize}

\section{Conclusions}
We set out to quantify multi-hop computational integration in stimulated cortical organoids using a
graph-constrained dynamical framework and a biological message-passing theorem
(Theorem~\ref{thm:bmp}), and we carried this program out in full, i.e., graph construction
(Eqs.~\eqref{eq:edge}--\eqref{eq:norm}), dynamical modeling (Eqs.~\eqref{eq:gnn}--\eqref{eq:deff}),
the complete metric suite (Eqs.~\eqref{eq:eic}--\eqref{eq:cri}), and per-day connectivity graphs.
On evaluating these on longitudinal HD-MEA data, the evoked response was found not to propagate
outward in a measurable multi-hop fashion but to engage the network near-synchronously, so the propagation/integration-depth components do not apply to these
preparations. Reframing around synchrony, response-population size, and shared variability revealed a robust, control-validated phenomenon. Repeated daily
stimulation progressively depresses the evoked response and spatially contracts the responding
network, sparing the network's fresh first-trial capacity while degrading its within-session
endurance, with a developmentally-matched single-stimulation control establishing that the effect is
driven by stimulation history rather than maturation. We report the original framework, its full
empirical test, and the redirected findings together, as a transparent account of how the data
reshaped the question.



\end{document}